\documentclass[useAMS,usenatbib,usegraphicx]{mn2e} 
\usepackage{graphicx} 
\graphicspath{{plots/}} 

\title[3.8-Micron Photometry of HD\,209458b]{3.8-Micron Photometry 
During the Secondary Eclipse of the Extrasolar Planet HD\,209458b}

\author[Deming et al.]{Drake Deming$^{1,4}$\thanks{E-mail:
ddeming@pop600.gsfc.nasa.gov}, L. Jeremy Richardson$^{2}$\thanks{E-mail:
richardsonlj@stars.gsfc.nasa.gov}, \& Joseph Harrington$^{3}$
\thanks{E-mail:jharring@physics.ucf.edu}\\
$^{1}$Planetary Systems Laboratory, Code 693, Goddard Space Flight
Center, Greenbelt MD 20771 USA\\
$^{2}$Exoplanet and Stellar Astrophysics Laboratory, Code 667, 
Goddard Space Flight Center, Greenbelt MD 20771 USA\\
$^{3}$Department of Physics, University of Central Florida, 
Orlando FL 32816-2385 USA\\
$^{4}$Visiting Astronomer at the Infrared Telescope Facility, which
is operated by the University of Hawaii\\ 
under Cooperative Agreement no. NCC~5-538 with the National 
Aeronautics and Space Administration,\\
Science Mission Directorate, Planetary Astronomy Program}

\begin{document}

\date{ }

\pagerange{\pageref{firstpage}--\pageref{lastpage}} \pubyear{2002}

\maketitle

\label{firstpage}

\begin{abstract}

We report infrared photometry of the extrasolar planet HD\,209458b
during the time of secondary eclipse (planet passing behind the star).
Observations were acquired during two secondary eclipses at the NASA
Infrared Telescope Facility (IRTF) in September 2003. We used a
circular variable filter (1.5\% bandpass) centered at 3.8 $\mu$m to
isolate the predicted flux peak of the planet at this
wavelength. Residual telluric absorption and instrument variations
were removed by offsetting the telescope to nearby bright comparison
stars at a high temporal cadence. Our results give a secondary eclipse
depth of $0.0013\pm0.0011$, not yet sufficient precision to detect the
eclipse, whose expected depth is $\sim 0.002 - 0.003$.  We here
elucidate the current observational limitations to this technique, and
discuss the approach needed to achieve detections of hot Jupiter
secondary eclipses at 3.8 $\mu$m from the ground.

\end{abstract}


\section{Introduction}

The passage of a transiting extrasolar planet behind its star during
secondary eclipse has emerged as a valuable technique for measuring
thermal radiation emitted by hot Jupiters. Secondary eclipse
detections of hot Jupiters have been accomplished using the Spitzer
Space Telescope \citep{charb05, deming05a, deming06a}. The spectra of
hot Jupiters are predicted to exhibit prominent flux peaks near 2.2 \&
3.8 $\mu$m, where absorption by water vapor is minimal
\citep{sudarsky, seager, fortney}.  Unfortunately, the bandpasses
available from Spitzer do not sample these flux peaks in an optimal
way. Spitzer has no capability at 2 $\mu$m, and the IRAC filter
bandpass at 3.5 and 4.5 $\mu$m \citep{fazio} overlap the 3.8 $\mu$m
peak minimally.

Observing hot Jupiter extrasolar planets at 2.2 and 3.8 $\mu$m in the
near future must depend on ground-based observations. Ground-based
observers have attempted both photometry \citep{snellen, snellen2} and
spectroscopy \citep{rich03} to detect the 2.2 $\mu$m peak using
secondary eclipses. Recently, a tentative detection of OGLE-TR-113 at
2.2 $\mu$m was reported by \citet{snellen2}, with eclipse amplitude
$0.0017\pm0.0005$.  \citet{knutson_a} attempted to detect the 3.8
$\mu$m peak (L-band) of TrES-1 spectroscopically, but no
investigations have yet reported photometry of hot Jupiters at 3.8
$\mu$m.  At this longer wavelength, thermal background emission from
the telescope and terrestrial atmosphere is a major impediment.
Moreover, observations in the L band are significantly affected by
telluric water vapor absorption, which is notoriously variable. In
addition, photometric transit searches at visible wavelengths have
been significantly affected by `red noise' \citep{pont}, and this
problem might become more severe in the infrared (IR). It is therefore
of interest to explore the limits of ground-based L-band photometry
for secondary eclipse detection.

In this paper we report L-band photometry during the secondary eclipse
of HD~209458b. A preliminary description of these measurements was
reported by \citet{deming06b}; here we report the final results.  We
do not achieve a detection of the planet's thermal emission, but we
elucidate the observational limitations, and we describe improvements
that will allow an L-band detection from the ground.

\section{Observations}

We observed HD~209458 during two secondary eclipses on 2003 September
9 \& 16 UT, using NASA's 3-metre Infrared Telescope Facility (IRTF) on
Mauna Kea.  We also observed (and detected) a primary eclipse
(transit) on 2003 August 17 UT.  Since the HD~209458b transit is well
observed at both visible \citep{brown} and IR wavelengths
\citep{rich06}, we concentrate here on the secondary eclipse
observations. However, observations on the transit night were very
useful in analyzing the properties of our data (see below).

We imaged HD~209458 using the 256x256-pixel \mbox{NSFCam} IR camera
\citep{shure}. To avoid detector saturation from thermal background
radiation, we used a circular-variable-filter (CVF) with a 1.5\%
bandpass tuned to 3.8 $\mu$m.  This wavelength not only corresponds to
a predicted peak in the planet's thermal emission, but also has optimal
transmission through the terrestrial atmosphere.  Figure~1 shows the
CVF bandpass in comparison to the telluric transmittance at Kitt Peak,
obtained by convolving the atmospheric transmission atlas of
\citet{atlas} to lower spectral resolution ($\lambda/\delta\lambda =
100$, comparable to CVF resolution).  Since Mauna Kea is higher and
drier than Kitt Peak, Figure~1 is a worst case representation of the
telluric transmission for our observations.
 
\begin{figure} 
   \resizebox{\hsize}{!}{ 
     \includegraphics[angle=0]{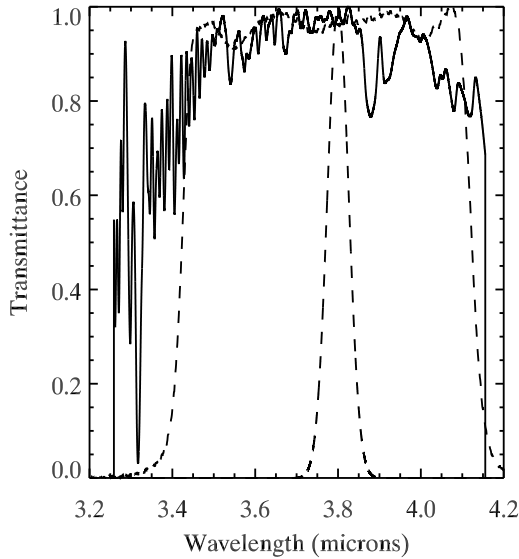}}     
   \caption{Telluric transmission near 3.8~$\mu$m, at a spectral
   resolution of $\lambda/\delta\lambda = 100$, based on the
   transmission atlas of \citet{atlas}, shown in comparison to the
   standard L$^\prime$ bandpass (wide dashed curve), and our CVF 
   bandpass (narrow dashed curve).} 
   \label{fig1} 
\end{figure} 
 
It is generally not feasible to monitor comparison stars
simultaneously with the target star in ground-based thermal IR
photometry. Attaining the requisite wide field of view would require
an increase in the per-pixel solid angle seen by the detector, to the
point where the observations would be unavoidably saturated by thermal
background. Instead, our observations alternated between HD~209458 and
a comparison star, moving the telescope between them at the most rapid
possible cadence (17 sec for telescope motion and settling). Our
imaging sequence consisted of a continuing series of exposures in the
order $T_a, C_a, C_b, T_b, T_a, ...$, where $T$ indicates the target
star (HD~209458), and $C$ indicates a comparison star (see below).
The subscripts $a$ and $b$ indicate two distinct nod positions on the
detector array, separated by 6 arcsec (110 pixels).  The nod is used
for subtraction of thermal background (see Infrared Photometry
section). The observations commenced each night before the time of
secondary eclipse, and continued for as long as possible after
eclipse.

The ideal comparison star would have an IR brightness comparable to
the target, and would lie at a close angular distance.  Unfortunately,
no ideal star exists near HD~209458.  We must choose between fainter
stars in close proximity, or brighter stars at greater angular
distances. Since these are exploratory observations, we elected to try
both approaches.  On 9~Sept we monitored HD~209346 (A2, $V=8.3$, $0.2$
degrees distant).  To compensate for this star being 0.7 magnitudes
fainter than HD~209458, we added additional comparison images to the
observing sequence, viz.: $T_a, C_a, C_b, C_a, C_b, T_b, T_a, ...$.
On 16~Sept we monitored HD~210483 (G1,V=7.6, 1.7 degrees distant),
following \citet{rich03}.  For all stars, images consisted of two
co-adds of 5 sec exposures. A single 5 sec exposure produced about
$2\times 10^8$ electrons of background radiation within our synthetic
aperture for photometry (see below), and about $5\times 10^6$
electrons due to HD~209458. 

\section{Infrared Photometry}

Because the thermal background is intense at this wavelength, weighting
the stellar image by the average point-spread-function (PSF) could in
principle achieve an optimum signal-to-noise ratio (SNR) for
photometry.  The stellar PSF at this wavelength exhibits a significant
component due to diffraction, as shown in Figure~2.  Unfortunately,
the PSF also contains a large contribution from seeing, and did not
prove to be sufficiently stable to utilize PSF-weighted photometry. We
therefore extracted photometric intensitites for the star using simple
aperture photometry, with a circular synthetic aperture of radius 20
pixels (1.1 arcsec). The best radius for the synthetic aperture was
determined by minimizing the scatter in the photometric intensities.
With radii significantly larger than 20 pixels, the greater background
noise within the aperture degrades the SNR, and with significantly
smaller radii the photometry becomes sensitive to seeing fluctuations.

\begin{figure} 
   \resizebox{\hsize}{!}{ 
     \includegraphics[angle=0]{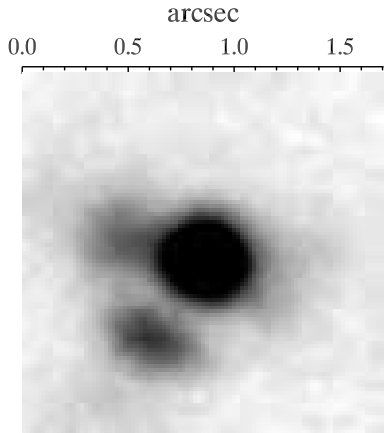}}     
   \caption{Background-subtracted image of HD~209458 at 3.8~$\mu$m, in
   reverse grayscale, showing telescope diffraction effects. The image
   core has been saturated to illustrate the weak diffraction wings,
   which are asymmetric due to optical imperfections. At this
   wavelength the PSF is a mixture of diffraction and seeing.}
   \label{fig2} 
\end{figure} 

A significant factor in the quality of our photometry is telescope
focus, which changes with temperature and has to be monitored and
corrected by the IRTF observer.  We found that both the background level
and the stellar intensity varied with telescope focus.  Figure~3
shows the background level for the HD~209458 transit observations
(best night to illustrate this effect). Improvements in telescope
focus are accompanied by a decrease in the measured background.  We
attribute this to less off-axis acceptance of warm radiation from the
telescope structure as focus improves. The dependence of background on
observed airmass allows us to deduce that $\sim 80\%$ of the background
originates from the telescope and warm optics, the remaining $\sim 20\%$
being contributed by the terrestrial atmosphere.
 
\begin{figure} 
   \resizebox{\hsize}{!}{ 
     \includegraphics[angle=0]{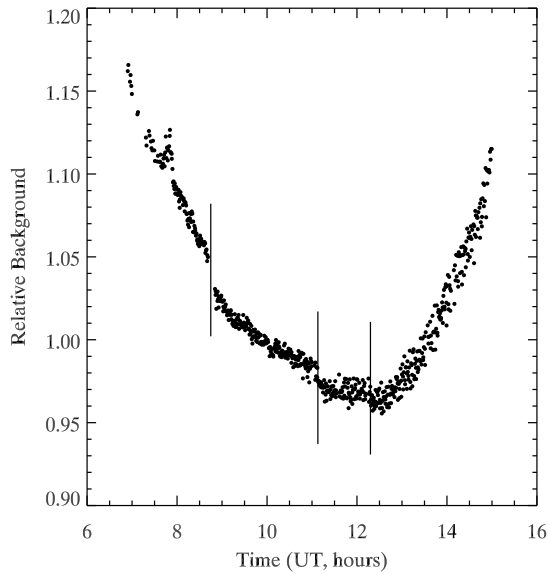}}     
   \caption{Thermal background intensity in our 3.8~$\mu$m bandpass,
   as observed on 17 August 2003.  The vertical lines indicate times
   of telescope re-focusing, showing that the background level
   decreases with updates in telescope focus.}
   \label{fig3} 
\end{figure} 

In principle, the stellar photometry should be independent of
telescope focus, as long as the synthetic aperture is of sufficient
radius to encompass slightly defocused stellar images. In practice, we
find that the stellar photometric values change by $\sim 1\%$ when
updating telescope focus. We found a weak correlation between
intensity and the width of the stellar PSF (correlation coefficient
$\sim 0.4$).  Broader stellar images - from poorer seeing - tended to
yield less intensity from aperture photometry.  Although this
correlation was dominated by seeing fluctuations, we expect that
imperfections in focus have a similar effect.  The sense of the
correlation - less photometric intensity from broader PSFs - is
consistent with not recovering all of the stellar photons within a
given synthetic aperture as the image broadens. Using extremely large
synthetic apertures is impractical due to background noise.  We
therefore analyzed our data in blocks between focus updates, measuring
HD~209458 relative to the comparison star.

Although thermal IR photometry is daunting in many respects, there are
some compensating advantages.  These include insensitivity to
intra-pixel variations, and the ability to flat-field the detector
array nearly simultaneously with the data acquisition.  Since the
thermal background will be spatially uniform over our $14$ arcsec
field of view, we use the thermal background for flat fielding.
Within each block defined by focus updates, we compute the median
image.  Since telescope nods and pointing jitter move the star by
significantly more than its FWHM, the median image does not contain
the star, and defines an accurate flat field calibration for that data
block.

We measure the intensity ratio between HD~209458 and the comparison
star, and we do this separately for the 'a' and 'b' nod positions in
each data block. To increase the precision of the comparison star
photometry, we smooth the individual comparison measurements using a
10-point moving average. This increases the effective time scale for
our HD~209458 to comparison ratio to $\sim$ 20 minutes.  We
spline-interpolate the comparison star moving average to the times of
the HD~209458 observations, and compute the ratio.  Results from this
process, for two representative data blocks, are shown in Figure~4. We
explored the possibility of improving the precision by decorrelating
the stellar intensities against the widths of their PSFs, but we found
that the correlation was not sufficiently strong to significantly
improve our results.

We compute errors for each photometric measurement of HD~209458. Both
the HD~209458 photometry and the comparison star photometry contribute
to the total error.  The largest source of error is background shot
noise, and we compute its magnitude by measuring the per-pixel
fluctuations in the flat-fielded background for each frame, using the
region immediately adjacent to the photometry aperture. This noise was
about two times larger than the theoretical value from the square root
of the electron number.  Scaling the measured fluctuation as the
square root of the number of pixels, we calculate the shot noise in
the synthetic aperture due to background.  Stellar photon noise and
detector read noise are negligible by comparison.  We propagate these
errors through the relative photometry, and compute the error in the
10-point moving average of the comparison stars.  Comparison star and
HD~209458 errors are added in quadrature, yielding an average
per-point fractional precision of 0.009 per single HD~209458
measurement. (In this paper we quote eclipse depths and errors in
units of the stellar intensity, {\it not} in magnitudes.)  This does
not include errors due to the terrestrial atmosphere, which may not
cancel perfectly between HD~209458 and the comparison star.
 
\begin{figure} 
   \resizebox{\hsize}{!}{ 
     \includegraphics[angle=0]{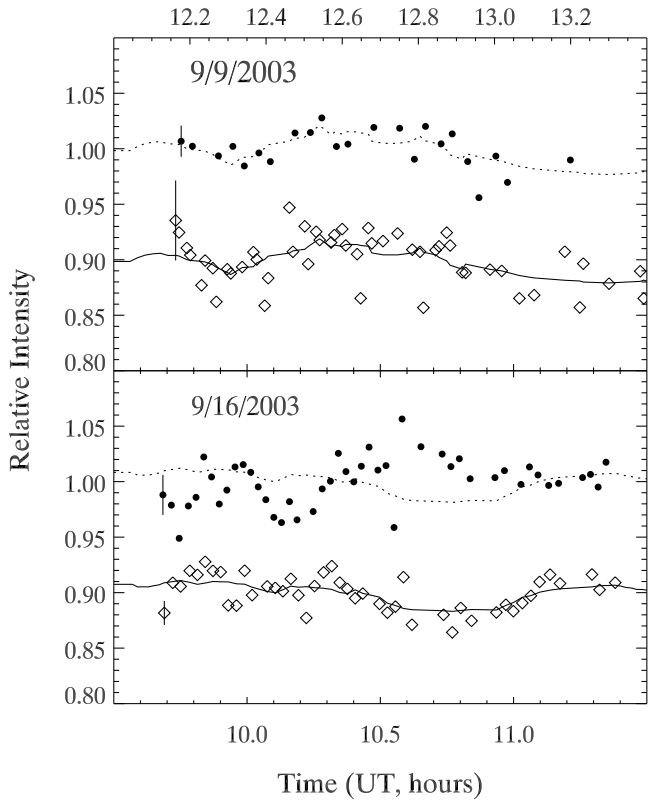}}     
   \caption{Sample 3.8~$\mu$m photometry UT 9 Sept 2003 (upper panel)
   and UT 16 Sept (lower panel). The open diamonds show observations
   of the comparison star (HD~209346 on 9 Sept \& HD~210483 on 16
   Sept). The comparison star points have been normalized to 90\% of
   HD~209458 (solid symbols), for clarity.  The solid line is a
   10-point moving average of the comparison star data; the dashed
   line scales this relation to HD~209458. Representative error bars
   (${\pm}2\sigma$ for clarity, and including only background
   fluctuations) are plotted over the first points in each
   series. Note the tendency for HD~209458 to track the nearby
   comparison star on 9 Sept, but not the more distant comparison star
   observed on 16 Sept.}
   \label{fig4} 
\end{figure}

\begin{figure} 
   \resizebox{\hsize}{!}{ 
     \includegraphics[angle=0]{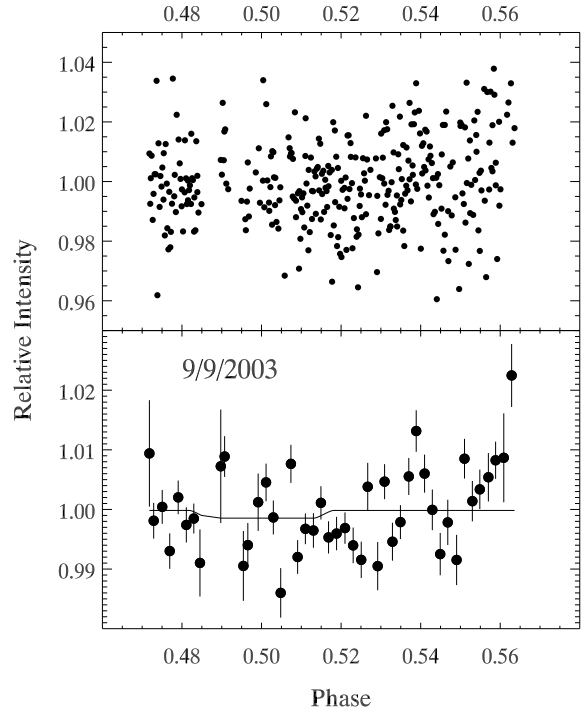}}     
   \caption{Results for 3.8~$\mu$m photometry of HD~209458b on UT 9
   Sept 2003.  Upper panel: 331 individual observations, shown
   without error bars for clarity.  Lower panel: results in bins of
   0.002 in phase, with error bars and the best-fit eclipse curve
   added.  The best-fit eclipse depth is $0.0013\pm0.0011$, not a
   significant detection.}
   \label{fig5} 
\end{figure} 

\section{Results and Discussion}

Although the photometry exhibits a general decrease in intensity with
increasing airmass, there are significant fluctuations on shorter time
scales. Measurements of comparison stars are necessary for this
reason. Examination of the photometry for the nights of 9 \& 16 Sept
shows much better results using the closer, albeit fainter, comparison
star (9 Sept).  This is obvious on Figure~4, where the lower panel (16
Sept) exhibits fluctuations on time scales of $\sim$ 20 minutes that
are not similar between HD~209458 and the comparison star. We find
that this holds on 16 Sept even when fewer points are used in the
moving average of this bright comparison star.  Therefore, the data on
this night are not useful for eclipse detection.  We conclude that the
relatively large angular separation between HD~209458 and HD~210483
(1.7 degrees) is the principal factor contributing to this significant
degree of atmospheric noise. Physically, this probably results from
incoherence of water vapor absorption over the distances separating
the lines of sight to these two stars. Note that this conclusion will
not necessarily apply to the results of \citet{rich03}, since those
authors used HD~210483 to perform {\it spectroscopy} relative to HD~209458.

In contrast to the 16~Sept results, the data for 9~Sept (e.g., upper
panel of Figure~4) show that fluctuations in HD~209458 consistently
track those of HD~209346, even though the effective time scale of
these data is $\sim$ 20 minutes (10-point moving average of the
comparison star). This is also true for the other data blocks on this
night. The increase in thermal background per unit airmass was
approximately two times lower on 9~Sept than 16~Sept (10\% vs. 20\%).
Nevertheless, we judge that the primary difference in our results is
due to the comparison star selection: a nearby, albeit faint,
comparison star is more useful than a brighter but more distant one
for photometry in the L-band.  Certainly this is true in conditions
that are less than optimum.

Figure~5 (upper panel) plots the 331 individual observations from
9~Sept versus the planet's orbital phase, using the ephemeris
from \citet{knutson_b}. These data are normalized to an average of
unity.  The lower panel averages the data in bins of width 0.002 in
phase, and adds error bars calculated as discussed above. The best-fit
eclipse depth ($0.0013\pm0.0011$) does not have sufficient precision
for detection, but our results are similar to a result recently
reported for TrES-1 by \citet{knutson_a}, who used a spectroscopic
technique at 3.8 $\mu$m.

The reduced chi-squared of the fit in the lower panel of Figure~5 is
1.96, consistent with a per-point scatter of 0.014 (upper panel),
about 50\% larger than the errors from background shot
noise. Additional error can arise from atmospheric noise that is not
common to both stars on the spatial and temporal scales of these
observations. This 'red noise' (amplitude $\sim$ 0.01) can be reduced
by using a broader optical bandwidth.  New IR cameras such as
\mbox{NSFCam-2} at the IRTF can utilize the full L' band without
saturating.  This will increase the SNR by a factor of $\sim 3.5$ per
image (SNR proportional to square root of optical bandwidth). Having
greater SNR per frame should increase the precision of fainter
comparison stars; for example, HD~209346 should improve from 0.021 to
0.006 precision in 10 seconds of integration. A more significant
effect of a broader optical bandwidth is that it will decrease the
temporal bandwidth of the red noise, by an order of magnitude.  This
follows because the time to define the target-to-comparison-star ratio
to a given precision shortens as the square of the SNR per image, and
we would not need to utilize a 10-point moving average for the
comparison star.

Atmospheric noise power often increases as the inverse of the sampling
frequency (`1/f' noise).  We have calculated the improvement expected
in both the background shot noise and the atmospheric 1/f noise using
the full L' optical bandwidth.  This calculation assumes that any high
frequency contribution from the terrestrial atmosphere - similar to
scintillation - does not dominate on $\sim 10$ to $30$ second time
scales at this wavelength.  The result indicates that our 9~Sept final
errors will decrease by a factor of two.  A secondary eclipse of depth
0.002 in HD~209458 would be detected to $3.6\sigma$ significance.
Moreover, the greater planet-to-star contrast of the recently
discovered HD~189733 system \citep{bouchy}, coupled with its brighter
apparent magnitude, will lead to even more favorable detectability.
We conclude that ground-based detection of photons from extrasolar
planets is possible at 3.8 $\mu$m.

\section*{Acknowledgments}

We thank our IRTF support scientist, Bobby Bus, for his expert help in
making the observations, and we are grateful to John Rayner for
assistance with NSFCam and to the IRTF telescope operators for their
assistance.

\label{lastpage}

\end{document}